\documentclass[useAMS,usenatbib]{mn2e}

\usepackage{epsfig}
\newcommand{\msun}{\mbox{$M_\odot$}}

\title[Galactic Outflows and Evolution of the ISM]
{Galactic Outflows and Evolution of the Interstellar Medium}
\author[Benoit C\^ot\'e et al.]
{Benoit C\^ot\'e,$^{1,2}$\thanks{E-mail: benoit.cote.4@ulaval.ca},
Hugo Martel,$^{1,2}$ Laurent Drissen,$^{1,2}$ and
Carmelle Robert$^{1,2}$\\
$^{1}$D\'epartement de physique, de g\'enie physique et d'optique,
Universit\'e Laval, Qu\'ebec, QC, G1V 0A6, Canada\\
$^{2}$Centre de Recherche en Astrophysique du Qu\'ebec}
\begin{document}

\date{Accepted XXX. Received XXX; in original form XXX}

\pagerange{\pageref{firstpage}--\pageref{lastpage}} \pubyear{XXX}

\maketitle

\label{firstpage}

\begin{abstract}
We present a model to self-consistently describe
the joint evolution of starburst galaxies and the galactic wind resulting from
this evolution. This model will eventually be used to provide a subgrid
treatment of galactic outflows in cosmological simulations of galaxy 
formation and the evolution of the intergalactic medium (IGM).
We combine the population synthesis code Starburst99 with
a semi-analytical model of galactic outflows and a model for the distribution
and abundances of chemical elements inside the outflows.
Starting with a galaxy mass, formation redshift,
and adopting a particular form for the star formation rate, we describe
the evolution of the stellar populations in the galaxy, the evolution of
the metallicity and chemical composition of the interstellar medium (ISM),
the propagation of the galactic wind, and the metal-enrichment
of the intergalactic medium. 
The model takes into account the full energetics of
the supernovae and stellar winds and their impact on the propagation
of the galactic wind, the depletion of the ISM by the galactic wind
and its impact on the subsequent evolution of the galaxy, as well as the 
evolving distributions and abundances of metals in the galactic wind.
In this paper, we study the properties of the model, by varying the 
mass of the galaxy, the star formation rate, and the efficiency of
star formation.
Our main results are the following:
(1) For a given star formation efficiency $f_*$, a more extended period
of active star formation tends to produce a galactic
wind that reaches a larger extent. 
If $f_*$ is sufficiently large, the energy deposited
by the stars completely expels the ISM. Eventually, 
the ISM is being replenished by mass loss from supernovae and stellar winds.
(2) For galaxies with masses
above $10^{11}\msun$, the material ejected in the IGM 
always falls back onto the galaxy. Hence
lower-mass galaxies are the ones responsible for enriching the IGM. 
(3) Stellar winds play a minor role in the
dynamical evolution of the galactic wind, because their energy input is
small compared to supernovae. However, they
contribute significantly to the chemical composition of the galactic wind. 
We conclude that the history of the ISM enrichment plays a determinant role
in the chemical composition and extent of the galactic wind, and therefore
its ability to enrich the IGM. 
\end{abstract}


\section{Introduction}

Galactic winds and outflows are the primary mechanism by which galaxies
deposit energy and metal-enriched gas into the intergalactic medium 
(IGM).\footnote{Some authors make a distinction 
between {\it galactic winds}, which are generated
over most of the lifetime of the galaxy and inject energy and
metals at a steady rate, and {\it galactic outflows}, which result from violent
processes like starbursts, are short-lived, and eject material
at large enough distances into the IGM to eventually reach other galaxies.
In this paper, we use one or the other to designate
any material that is ejected from the galaxy and deposited into the IGM.}
This can greatly affect the evolution of the IGM, and the subsequent
formation of other generations of galaxies. 
Feedback by galactic outflows can provide an explanation for
the observed high mass-to-light ratio of dwarf galaxies and the
abundance of dwarf galaxies in the Local Group,
and can solve various problems with galaxy formation models, such as
the overcooling and angular momentum problems
(see \citealt{benson10} and references therein).
Galactic outflows can explain
the metals observed in the IGM via the Lyman-$\alpha$ forest
(e.g. \citealt{my87,schayeetal03,ph04,aguirreetal08,pierietal10a,
pierietal10b}),
the entropy content and scaling relations in X-ray clusters
\citep{kaiser91,eh91,cmt97,tn01,babuletal02,voitetal02},
and provide observational tests that can constrain 
theoretical models of galaxy evolution. Local examples of spectacular 
outflows in dwarf starburst galaxies include those of the extremely 
metal-poor I Zw 18 (\citealt{pequi08,jamet10}) and NGC1569 \citep{west09}.
More massive spirals, such as NGC7213 \citep{hameed01}, also show 
evidence of global outflows. For a review of the subject, see \citet{vcbh05}.

\subsection{Galactic Outflow Models}

Large-scale cosmological simulations have become a major tool in the
study of galaxy formation and the evolution of the IGM at
cosmological scales. These simulations start at high redshift
with a primordial mixture of dark and baryonic matter, and a spectrum
of primordial density
perturbations. The algorithm simulates the evolution of the 
system by solving the equations of gravity, hydrodynamics, and 
(sometimes) radiative 
transfer. Adding the effect of galactic outflows in these
simulations poses a major practical problem. In one hand,
the computational volume
must be sufficiently large to contain a ``fair'' sample of the universe,
typically several tens of Megaparsecs. On the other hand, the 
physical processes responsible for generating the outflows take place
inside galaxies, at scales of kiloparsecs or less. This represents at
the very minimum 4 orders of magnitude in length and 12 orders of magnitude
in mass, which is beyond the capability of current computers. 
Since we cannot simulate both large and small scales simultaneously,
the usual solution consists of simulating the larger scales and using 
a {\it subgrid physics} treatment for the smaller scale.
Cosmological simulations can predict the location of the galaxies
that will produce the outflow, but cannot resolve the inner structure of
these galaxies with sufficient resolution to simulate the actual
generation of the outflow. Instead, the algorithm will use a prescription
to describe the propagation of the outflow and its effect on the 
surrounding material.

One possible approach
consists of depositing momentum or thermal energy ``by hand'' 
into the system, to simulate the effect of galactic outflows on
the surrounding material \citep{std01,theunsetal02,sh03,cno05,od06,
kollmeieretal06}. The algorithm determines the location of the galaxies
producing the outflows and calculates the amount of momentum or thermal
energy deposited into the IGM based on the galaxy properties
(mass, formation redshift $\ldots$). Then, in particle-based algorithms like
smoothed particle hydrodynamics
(SPH), this momentum or energy is deposited on the nearby particles, while
in grid-based algorithms it is deposited on the neighboring grid points.
This will result in the formation and expansion of a cavity around each 
galaxy, which is properly simulated by the algorithm.

A second approach consists of combining the numerical
simulation with an analytical model for the outflows.
Tegmark et al. (1993) have developed an analytical model to
describe the propagation of galactic outflows in an expanding universe.
In this model, a certain amount of energy is released into the
interstellar medium
(ISM) by supernovae (SNe) during an initial starburst. This energy drives
the expansion of a spherical shell that propagates into the surrounding
IGM, until it reaches pressure equilibrium. This
model, or variations of it, has 
been used extensively to study the effect of galactic outflows
on the IGM (\citealt{fl01,mfr01,sb01,sfm02,sh04,lg05,pmg07},
hereafter PMG07; \citealt{sss08,gbm09,pmp10}).
In this approach, the evolution of the IGM and the propagation of the outflow
are calculated 
separately, but not independently as they can influence one another.
The presence of density inhomogeneities in the IGM can affect 
the propagation of the outflow, while energy and metals carried by the
outflow can modify the evolution of the IGM.

There are several limitations with this second approach.
In particular, it assumes that the initial starburst, which occurred
during the formation of the galaxy,
is the only source of energy driving the expansion of the outflow. First,
the starburst lasts for a short period of time, typically 50 million years. 
Hence, we would expect to observe very few galaxies having an outflow.
Furthermore these galaxies would be just forming and therefore would
have complex and chaotic structures. Observations show 
instead that outflows are ubiquitous and often originate from 
well-relaxed galaxies (see \citealt{vcbh05} and references therein).
Second, even though the injection rate of energy is maximum during
the initial starburst, the total amount of energy which is injected afterward
by all generations of SNe and stellar winds
could be comparable or even more important. Even if this energy is
injected slowly over a large period of time, the cumulative effect
could be significant. Indeed, an initial outflow caused by
a starburst could be followed by a steady galactic wind that 
would last up to the present. The role of stellar winds
has been mostly ignored in analytical models and numerical simulations of
galactic outflows. Third, there is the possibility that accretion of 
intergalactic gas onto the galaxy might trigger a second starburst.
Finally, a recent study \citep{sbh10} suggests
that there is a significant time delay between the initial starburst
and the onset of the outflow, something not considered by current models.

Another important issue is the amount of metals contained in the
outflow, the spatial distribution of metals in the outflow, and the relative
abundances of the various elements.
The metallicity
of the outflow depends on the metallicity of the ISM at the time of
the starburst. The metals contained in the ISM at that time 
can have several origins: (1) metals already present in the gas when the 
galaxy formed, (2) the SNe produced during the starburst, and
(3) the stellar winds generated by massive stars and AGB objects.
Hence, the composition of the outflows will depend on the epoch of
formation of the galaxy (which determines the initial metal abundances),
as well as the relative amount of metals injected into
the ISM by Type~Ia SNe, core-collapse SNe (Types~Ib, Ic and II), and winds,
and the timing of these various processes.
As for the amount of metals injected in the IGM, 
models often assume that it is proportional
to the mass of the galaxy, and 
do not provide a description of the distribution of metals in the outflow
and the relative abundances of the elements (\citealt{sb01}; PMG07).

\subsection{Objectives}

{\it Our goal is to develop a new galaxy evolution model to improve
the treatment of galactic winds in cosmological simulations.}
This model will describe not only an initial starburst and its resulting
outflow, but the entire subsequent evolution of a galaxy
up to the present. It will take
into account the progressive injection of energy by SNe and stellar
winds (which could cause a steady galactic wind that would follow the
outflow and last up to the present), 
and the time-evolution of the metallicity and composition of the ISM
that would directly affect the composition of the galactic wind.
It will also provide a description of the metal content, metal distribution,
and chemical composition of the galactic wind. This emphasis on the 
structure and composition of the galactic
wind is what distinguish our model from recent semi-analytical models of
galaxy formation, which tend to focus on reproducing the properties of
the galaxies themselves (luminosity and mass functions, 
halo properties, disk sizes, $\ldots$).
For a review of the various semi-analytical models, see \citet{baugh06}.
Our approach combines a population synthesis algorithm to describe the stellar
content of the galaxy, an analytical model for the expansion of the galactic
wind, and a new model for the distribution of elements inside the
galactic wind.

The paper is organized as follows: in \S2, we
describe the method we use to calculate the mass and chemical composition
of stellar winds and
SN ejecta, a key ingredient of our algorithm. In \S3, we describe
the basic equations for the evolution of the ISM. In \S4, we describe our
galaxy wind model. Results are presented in \S5, and conclusions
in \S6.

\section{POPULATION SYNTHESIS MODEL}

\subsection{Starburst99}

Starburst99 \citep{lrd92,lh95,leithereretal99,vl05} 
is the population synthesis code
we selected to simulate the properties of the stellar winds
and SNe produced by stellar populations with various characteristics.
In this project, all simulations use an instantaneous
SFR to produce a stellar population of mass $M_{\rm pop}=10^6\,\msun$,
with a standard IMF as defined by \citet{kroupa01}. This population
is evolved up to a final time~$t_{\rm final}=1{\rm Gyr}$, 
with a timestep $\Delta t=10^4{\rm yr}$ (small enough to reproduce
each evolutionary phase).
The evolutionary tracks are a combination of the Geneva tracks at young 
ages ($\leq10^8{\rm Myr}$) with the Padova tracks at old ages, to better
reproduce the mass loss phases of the massive stars as well as those of
smaller-mass objects.
We consider four values of the initial metallicity 
($Z_i=0.001$, 0.004, 0.008, and solar metallicity 0.02).
We also specify that all stars with an initial mass $M_i$ of 8 to $120\,\msun$
will end as SNe. Binary systems which may also produce SNe are not included in
this code and not considered here.

The mass we chose for the stellar population, $M_{\rm pop}=10^6\,\msun$,
is large enough to provide a good sampling of the IMF. Yet, it is much
smaller than the stellar mass of even dwarf galaxies. This enables us to
simulate any galaxy SFR we want. To simulate an instantaneous starburst
with a stellar mass larger than $M_{\rm pop}$, 
we simply rescale the results
produced by Starburst99. To simulate a SFR that is extended over 
a finite period
of time, we offset the simulations, such that during each timestep $\Delta t$,
the correct stellar mass is formed. In \S\ref{subsec_initial} below,
we describe the various SFRs considered in this study.

\begin{figure}
\begin{center}
\vskip5pt
\includegraphics[width=3.3in]{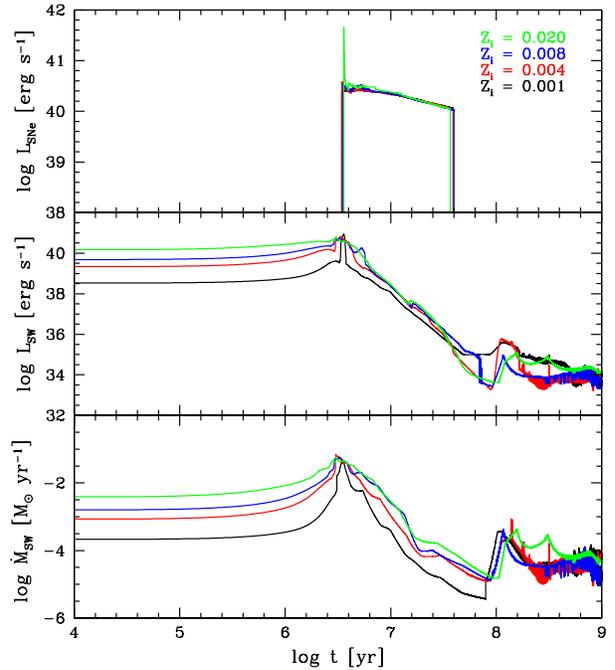}
\caption{Energy and mass ejected by a single stellar population,
versus time.
The SNe mechanical energy ($L_{\rm SNe}$, top panel),
stellar wind mechanical energy ($L_{\rm SW}$, middle panel),
and mass loss rate by stellar winds ($\dot M_{\rm SW}$, bottom panel)
have been obtained for a $10^6\msun$ stellar population, using Starburst99.
Different metallicities have been considered as indicated in the top panel.}
\label{lum_SN}
\end{center}
\end{figure}

Starburst99 provides, as a function of time, the mechanical energy produced
by SNe and stellar winds, the mass loss by
stellar wind, and the chemical composition of the stellar winds,
which takes into account the following elements:
H, He, C, N, O, Mg, Si, S, and Fe.
The code can also predict
the mass loss by SNe, but for this project we prefer to use a more detailed
treatment which also gives the chemical abundances of the SN yields
for a more extensive list of elements.
This treatment is described in the next section. Therefore, the ejected mass
calculated by Starburst99 will only include the contribution of stellar
winds. In Figure~\ref{lum_SN}, we plot the luminosity
(rate of mechanical energy injection) associated to SNe,
$L_{\rm SNe}$, and stellar winds, $L_{\rm SW}$, and
the mass loss by stellar winds, $\dot M_{\rm SW}$, 
versus time, for
stellar populations with different metallicities. The initial metallicity
has very little effect on the SNe luminosity, but
greatly impacts the luminosity and mass loss by stellar winds.
This implies that stellar
populations that form later will have more powerful winds, since they form out
of an ISM that has already been enriched in metals by earlier populations.
Figure~\ref{lum_SN} shows a sudden increase in
wind power around 1 - $2\times10^6{\rm yr}$, just prior to the arrival
of the first SNe, which is caused by the evolved stages of OB
stars, i.e. the Wolf-Rayet stars. 
The increase in mass loss after $10^8{\rm yr}$ is caused by the
low-mass stars on the AGB.

\subsection{SN Abundances}

Except in situations where the metallicity of the stars is larger
or equal to solar,
the enrichment of the ISM is dominated by SNe (see Fig.~8 of \citealt{lrd92}).
Thus, it is important to know the composition of the material ejected
by SNe as a function of the initial mass and metallicity of the
stellar progenitors. Several groups have used models of stellar evolution
and nucleosynthesis to produce tables of chemical abundances for SN ejecta.
These studies are summarized in Table~\ref{table_SNe}. 
The first column gives the authors of the papers;
the second and third 
columns list the metallicities and initial masses considered, respectively;
the fourth column indicates whether or not mass loss prior to the SN
explosion was included in the model.

\begin{table*}
 \centering
 \begin{minipage}{140mm}
  \caption{SN yields available in the literature.}
  \begin{tabular}{@{}llll@{}}
  \hline
   Paper & $Z$ & $M_i\>[\msun]$ & $\dot M$ \\
 \hline
 \citet{ww95} & 0, $2\times10^{-6}$, $2\times10^{-4}$, 0.002, 0.02 & 
      11, 12, 13, 15, 18, 19, 20, 25, 30, 35, 40 & no \\
 \citet{cl04} & 0, $10^{-6}$, $10^{-4}$, 0.001, 0.006, 0.02 
              & 13, 15, 20, 25, 30, 35 & no \\
 \citet{nomotoetal06} &
0, 0.001, 0.004, 0.02 & 13, 15, 18, 20, 25, 30, 35, 40 & yes \\
 \citet{wh07} & 0.02 &  32 $M_i$'s between 12 and 120 & yes \\
 \citet{lc07} & 0.02 &  15 $M_i$'s between 10 and 120 & yes \\
 \citet{hw10} & 0    & 120 $M_i$'s between 10 and 100 & no \\
\hline
\end{tabular}
\label{table_SNe}
\end{minipage}
\end{table*}

Although very useful,
these studies are not fully satisfactory for our work. The first three 
studies (\citealt{ww95,cl04,nomotoetal06}, hereafter N06) 
consider a wide range of metallicities,
from no metallicity ($Z=0$) to solar metallicity ($Z_\odot=0.02$), 
but are limited to initial masses $M_i\leq40\,\msun$.
The next three studies (\citealt{wh07}, hereafter WH07; \citealt{lc07};
\citealt{hw10}, hereafter HW10)
consider initial masses up to 100 - $120\,\msun$,
but only one value of the metallicity. Also, three of these studies do not
include mass loss prior to the explosion, 
which is critical in our models since we include
stellar wind effects. The study that comes the closest to our needs is
the one of N06. This is the only study that covers 
metallicities in the range $Z=0$ - $Z_\odot$ and includes mass loss. However,
its mass range only goes from 13 to $40\,\msun$.
These tables need to be extrapolated both at the low-mass and high-mass end 
to cover the full range of SNe progenitor masses.
In this paper, we assume an IMF with lower and upper mass limits of 
$1$ and $120\,\msun$, respectively, and SNe progenitor masses
in the range $M_i=8$ - $120\,\msun$. The lower limit of
$8\msun$ is the most commonly used, but that value is actually
quite uncertain, and could have an important effect on the results
(see Fig.~\ref{masseject_pop} below).

\begin{figure}
\begin{center}
\includegraphics[width=3.3in]{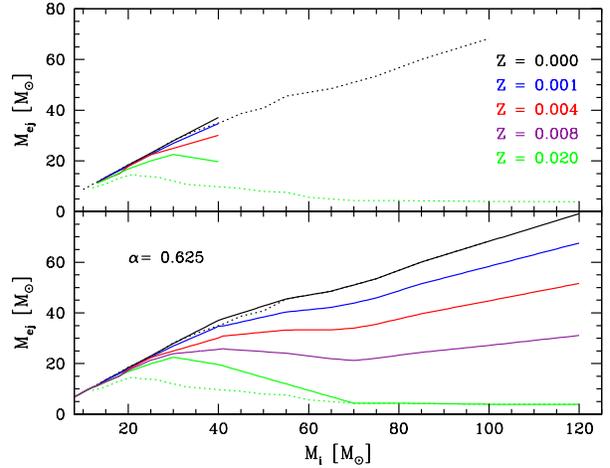}
\caption{Top panel: Total mass ejected by one SN versus the initial mass
of the progenitor, according to the tables of N06 (solid lines),
WH07 (lower dotted line), and HW10 (upper dotted line).
Colors represent various metallicities, as indicated.
Bottom panel: Total mass ejected by one SN versus the initial mass
of the progenitor
after extrapolating down to $M_i=8\,\msun$ and up to $M_i=120\,\msun$,
for metallicities $Z=0$ (solid black line) and $Z=Z_\odot$ 
(solid green line), after extrapolating over
all metallicities available in N06 (blue and red lines), 
and after interpolating to the metallicity $Z=0.008$ used by Starburst99
(purple line).
}
\label{masseject}
\end{center}
\end{figure}

The top panel of
Figure~\ref{masseject} shows the mass ejected by SNe 
versus initial mass and metallicity, calculated 
with the tables of N06 (solid curves). 
An interesting feature is the linear relation between $M_{\rm ej}$ and
$M_i$ at $Z=0$. Also, all the curves converge together at $M_i=13\,\msun$, 
suggesting that for initial masses $M_i<13\,\msun$, the
ejected mass is independent of metallicity. To extrapolate down to
$M_i=8\,\msun$, we used
the composition for $M_i=13\,\msun$, modulated using the
linear regression at $Z=0$ calculated with the software Slope
\citep{isobeetal90}. The result is also shown in Figure~\ref{masseject}.

Extrapolation up to $120\,\msun$ can be tricky, 
especially if mass loss is considered.
Several approaches have been suggested in the literature, especially at the
high-mass end. 
\citet{martinezserranoetal08}
extrapolate the mass ejected $M_{\rm ej}$ and chemical
composition of the ejecta linearly up to $M_i=100\,\msun$.
\citet{od08} assume similar ejecta for stars in the range
$M_i=35$ - $100\,\msun$ and also for stars in the range
$M_i=10$ - $13\,\msun$.
\citet{tornatoreetal07} and \citet{scannapiecoetal05}
assume that stars with initial masses larger than
$40\,\msun$ end up directly in a black hole, with no ejecta.
Using linear interpolation up to $M_i=120\,\msun$ would be risky. 
The relation between $M_{\rm ej}$ and
$M_i$ becomes metallicity-dependent for $M_i>25$ - $30\,\msun$, implying
that only two or three points would be available to extrapolate over
a mass interval three times wider than the one of N06. Furthermore
at $Z=Z_\odot$, $M_{\rm ej}$ eventually decreases with
increasing initial mass and a linear extrapolation would eventually lead
negative values before reaching $M_i=120\,\msun$. Setting the ejected mass 
to zero for $M_i>40\,\msun$
is not a good solution either. Even if the
SN results in the formation of a black hole, that black hole is never
as massive as the progenitor (\citealt{whw02}; WH07; 
\citealt{lc08,zwh08,belczynskietal10}). 
Using a copy of the $M=40\,\msun$ at larger masses 
is not realistic either since the tendency clearly shows that there is
more material ejected for larger initial masses, except at $Z_\odot$.

Since none of the three extrapolation methods considered so far
is satisfactory for extrapolating up to $120\,\msun$, we developed 
an alternative method. We started by extrapolating the table of N06
at $Z=0$ and also $Z_\odot$, by combining them
with the tables of HW10 and WH07, respectively.
The top panel of Figure~\ref{masseject}
shows
the $Z_\odot$ relation of WH07 and the $Z=0$ relation of HW10
(dotted curves). The 
$Z=Z_\odot$ relations of N06 and WH07 differ significantly in the range
$M_i=20$ - $40\,\msun$. This is likely the consequence of using a different
prescription for the mass loss during the pre-SN phase. 
In the model of WH07, the ejected mass reaches a maximum around
$M_i=21\,\msun$, then decreases with increasing initial mass, and finally
reaches a plateau at $M_{\rm ej}=4\,\msun$, which represents the minimum
mass a SN can eject. Assuming that this limit is correct, we
extrapolated the $Z_\odot$ model of N06 until it reaches the model of WH07
(around $M_i=70\,\msun$). We then switched to the model of WH07 to complete
the extrapolation up to $M_i=120\,\msun$. For $Z=0$, the results of N06
and HW10 are essentially identical in the range $M_i=13$ - $30\,\msun$,
and very similar in the range $M_i=30$ - $40\msun$. To combine them, we used
the tables of N06 in the range $M_i=8$ - $40\,\msun$ and 
the tables of HW10 in the range $M_i=55$ - $120\,\msun$,
Between 40 and $55\,\msun$, we interpolated between these two masses.
The results are shown in the bottom panel
of Figure~\ref{masseject}.

At the two intermediate metallicities $Z=0.001$ and 0.004, 
the values of the ejected masses should lie 
between the values for $Z=0$ and $Z_\odot$ (Fig.~\ref{masseject}). 
If the mass loss due to
stellar winds were neglected, the masses ejected by SNe should
be the same as for models with $Z=0$. The mass loss associated with
massive stars is proportional to $Z^\alpha$, where $\alpha\simeq0.6$ - 0.8
\citep{vdkl01,vdk05,krticka06,mokiemetal07}. We then interpolated
to get the mass ejected by SNe at any metallicity:
\begin{equation}
M_{\rm ej}(M_i,Z)=M_{\rm ej}(M_i,0)-\tau(M_i,Z)\dot M\,,
\quad\dot M=AZ^\alpha\,,
\label{mej1}
\end{equation}

\noindent where $A$ is a constant, and $\tau(M_i,Z)$ is the lifetime
of a star of initial mass $M_i$ and metallicity $Z$.
We made the approximation that the lifetime does not vary much with
metallicity, $\tau(M_i,Z)\approx\tau(M_i)$. Equation~(\ref{mej1}) becomes:
\begin{equation}
M_{\rm ej}(M_i,Z)=M_{\rm ej}(M_i,0)-B(M_i)Z^\alpha\,.
\label{mej2}
\end{equation}

\noindent where the function $B(M_i)$ was calculated from the 
masses ejected at $Z=Z_\odot$:
\begin{equation}
B(M_i)={M_{\rm ej}(M_i,Z_\odot)-M_{\rm ej}(M_i,0)\over Z_\odot^\alpha}\,.
\label{bmi}
\end{equation}

\noindent We choose the value $\alpha=0.625$, which minimizes 
the discontinuity at $M=40\,\msun$. Figure~\ref{masseject} shows all
the models of N06 after extrapolation. For $Z=0.004$ and 0.001,
the mass loss is not as extreme as in the case $Z=Z_\odot$. For that
reason, we used the composition of the $40\,\msun$ models modulated according
to the value of $M_{\rm ej}$ for all the initial masses $M_i>40\,\msun$.

\begin{figure}
\begin{center}
\includegraphics[width=3.3in]{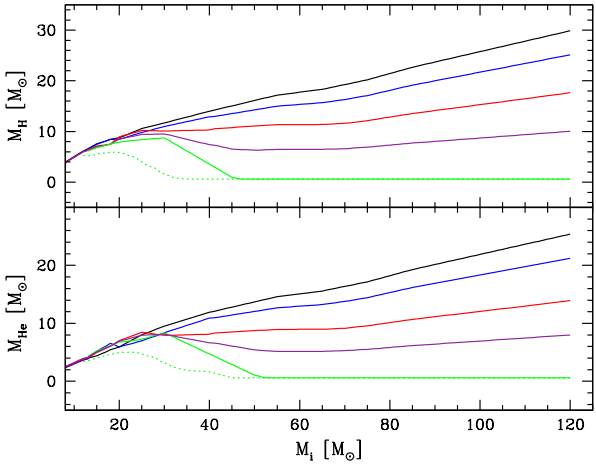}
\caption{Hydrogen (top) and helium mass (bottom)
ejected by one SN versus the initial mass
of the progenitor, according to the tables of N06 (solid lines),
after extrapolating down to $M_i=8\msun$ and up to $M_i=120\msun$, 
for $Z=Z_\odot$. The dotted line
illustrates the $Z=Z_\odot$ model of WH07. Colors have 
the same meaning as in 
Figure~\ref{masseject}.}
\label{masseject_H}
\end{center}
\end{figure}

Next we determined the composition of the ejected masses,
for $Z=0$ and $Z_\odot$.
Table~\ref{table_masseject} shows the hydrogen mass, helium mass,
and total mass ejected for the WH07 models. The cases presented are extreme.
The mass loss by stellar winds is sufficiently large to expel the entire
hydrogen envelop and most of the helium as well. This indicates that these 
stars have evolved through a Wolf-Rayet stage, resulting in SNe of Type~Ib
or Ic. Hence, most of the material ejected is composed of metals. For this 
reason, we treated hydrogen and helium as we treated the total mass 
ejected: we followed the tables of N06 until it reaches the model of WH07
and then switched to the model of WH07 (Fig.~\ref{masseject_H}). 
The reminder of the ejected
mass was assumed to have the chemical composition corresponding to the
$M_i=40\,\msun$ model. For $Z=0$, the complete composition of the 
$M_i=40\,\msun$ model was used.

\begin{table}
 \centering
 \begin{minipage}{80mm}
  \caption{Hydrogen mass, helium mass, and total mass ejected by SNe
    versus initial mass at solar metallicity, according to the table of WH07.}
  \begin{tabular}{@{}cccc@{}}
  \hline
   $M_i[\msun]$ & $M_{\rm H}[\msun]$ & $M_{\rm He}[\msun]$ 
 & $M_{\rm ej}[\msun]$ \\
 \hline
40  & 0.223               & 1.640 & 9.74 \\
50  & $5.51\times10^{-8}$ & 0.068 & 7.94 \\
60  & $1.77\times10^{-7}$ & 0.120 & 5.65 \\
70  & $1.16\times10^{-7}$ & 0.152 & 4.35 \\
80  & $1.04\times10^{-7}$ & 0.156 & 4.34 \\
100 & $1.77\times10^{-7}$ & 0.181 & 3.96 \\
120 & $1.39\times10^{-7}$ & 0.172 & 3.92 \\
\hline
\end{tabular}
\label{table_masseject}
\end{minipage}
\end{table}

Starburst99 requires tables of SNe at metallicities $Z= 0.001$, 0.004,
0.008, and 0.02. At this point we have tables at $Z=0.004$ and $0.02$.
For $Z=0.008$, we interpolated between the tables at $Z=0.004$
and 0.02
using $\log Z$ as an interpolation variable. 
We now have a full set of SN tables that provides the ejected mass and 
composition of the ejecta, for all progenitors in the range 
$M_i=8$ - $120\,\msun$ and $Z=0$ - $Z_\odot$.
Figure~\ref{masseject}
shows the ejected mass versus progenitor mass.

\begin{figure}
\begin{center}
\includegraphics[width=3.3in]{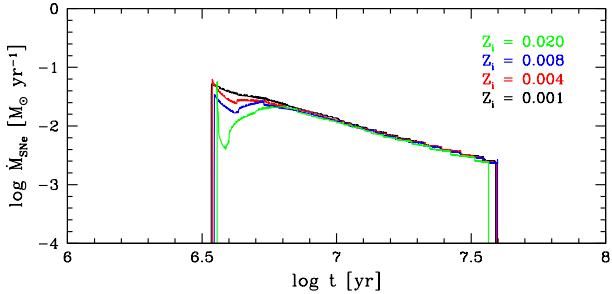}
\caption{Mass loss rate by SNe versus time.
Number of stars and SN explosions are from a Starburst99 model for
an instantaneous burst with a standard IMF and a total mass of
$10^6\msun$. Mass ejected by each
SN comes from our extended tables.
Different metallicities have been considered as indicated.}
\label{massloss_SN}
\end{center}
\end{figure}

\begin{figure}
\begin{center}
\includegraphics[width=3.3in]{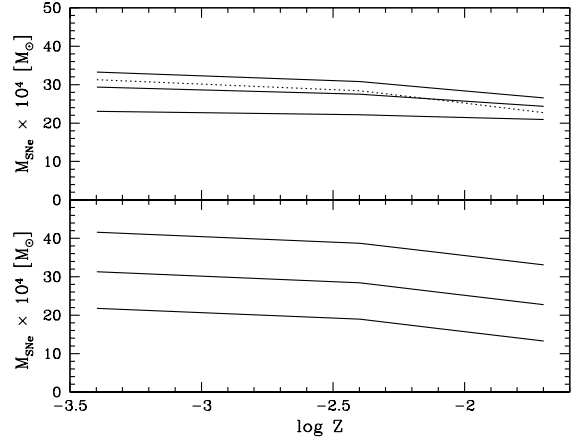}
\caption{Mass ejected by SNe versus metallicity,
Top panel: 
results obtained with our extrapolation method for the SN mass ejected as a
function of the progenitor initial mass, using the same stellar
population model as in Figure~\ref{massloss_SN} (dotted curve),
compared with results obtained with different
extrapolation methods (solid curves):
from top to bottom: linear extrapolation of \citet{martinezserranoetal08}, 
constant approximation of \citet{od08}, and no-ejecta approximation
of \citet{tornatoreetal07} and \citet{scannapiecoetal05}.
Bottom panel results obtained with our extrapolation method, for various
values of the minimum initial mass. From top to bottom,
$M_i=6\msun$, $8\msun$, and $10\msun$.
}
\label{masseject_pop}
\end{center}
\end{figure}

At each timestep during the simulation, Starburst99 provides the number
of SNe and their luminosity (adopting that each SN produces 
$10^{51}{\rm ergs}$; \citealt{mckee90}).
The SN tables are then used to calculate at every
timestep the mass and composition of the material deposited into the ISM.
For simulations with initial metallicity $Z=0$, we use the luminosity at 
$Z=0.001$ which is the smallest value considered by Starburst99.
Figure~\ref{massloss_SN} shows the mass loss rate 
due to SNe for a $10^6\,\msun$ instantaneous burst.
This completes the results shown in Figure~\ref{lum_SN}. 
In the top panel of Figure~\ref{masseject_pop}, we plot the total
mass ejected by SNe versus metallicity, for the same population.
The dotted curve shows our extrapolation method, while
the solid curves show the three other methods.
The ``constant'' method of \citet{od08} (using the
ejecta for $M_i=40\msun$ at higher masses) is the one which resembles 
our method the most, but there are significant differences. This justifies
{\it a posteriori} the method we have developed in this section.

We assume a minimum progenitor mass of $M_i=8\msun$, but we
also generated other tables using different values.
The bottom panel of Figure~\ref{masseject_pop} shows the ejected mass
for minimum progenitor masses of $M_i=6\msun$, $M_i=8\msun$ (the value
assumed in this paper), and $M_i=10\msun$. The differences are of order 50\%,
with the ejected mass being larger for a smaller minimum progenitor mass.
This shows that the particular
choice of minimum progenitor mass can have a impact
on the results, and we intend to investigate this issue in more details
in future work. Here we focus on the case of a minimum progenitor
mass of $M_i=8\msun$.

In a recent paper, \citet{horiuchi11} point to a serious ``supernova rate 
problem'': the measured cosmic massive-star formation
rate predicts a rate of core-collapse supernovae about twice as large as 
the observed rate, at least for redshifts between 0 and 1,
where surveys are thought to be quite complete. Several explanations are 
proposed to explain this major discrepancy, including
a large fraction of unusually faint (intrinsically or dust-attenuated), 
and thus unaccounted for, core-collapse SNe and a possible
overestimate of the star formation rate based on the current estimators.
If indeed this supernova rate problem is real, the
SFR (see \S5.1) might have to be scaled accordingly.
However, all simulations presented in this paper start at redshift
$z=15$, and terminate at redshifts between 6 and 9. It is not clear that
there is a supernova rate problem at these redshifts.

\section{EVOLUTION OF THE INTERSTELLAR MEDIUM}

During the evolution of a galaxy, the ISM is constantly enriched by
ejecta from stellar winds and SNe. Hence, every generation
of stars provides an environment richer in metals for the
future generations. The level of this enrichment depends on the SFR,
since the metal production increases with the number of stars formed.
To simulate this process, we designed an algorithm that combines 
the outputs of Starburst99 with the SNe tables of N06.

\subsection{Initial Conditions}
\label{subsec_initial}

We consider a galaxy with a total mass $M_{\rm gal}$. We assume
that the ratio of baryons to dark matter in the galaxy is equal to the
universal ratio, which is a valid assumption for the initial stages
of the galaxy. The baryonic mass of the galaxy is then given by 
\begin{equation}
M_b={\Omega_{b0}\over\Omega_0}M_{\rm gal}\,,
\label{mbaryons}
\end{equation}

\noindent where $\Omega_0$ and $\Omega_{b0}$ are the total and 
baryon density parameters, respectively. We assume that each galaxy starts
up with a primordial composition of hydrogen and helium
($X=0.755$ and $Y=0.245$).
The formation of the galaxy
results in a starburst, during which a fraction $f_*$ of the baryonic
mass is converted into stars. The total mass in stars at the end of
the starburst is therefore
\begin{equation}
M_*=f_*M_b\,.
\label{mstar}
\end{equation}

\noindent We refer to the parameter $f_*$ as the star formation efficiency.

We consider three different types of star formation rate: 
instantaneous, constant, and exponential. With an instantaneous SFR,
all stars form at $t=0$. In the other cases, the stellar mass formed $M_*$
and the star formation rate $\dot M_*$ are related by
\begin{equation}
\int_0^{t_f}\dot M_*(t)dt=M_*\,,
\end{equation}

\noindent where $t_f$ is the final time of the simulation. For a 
constant SFR, we have
\begin{equation}
\dot M_*={M_*\over t_{\rm burst}}\,,
\label{SFR_const}
\end{equation}

\noindent where $t_{\rm burst}$ is the duration of the starburst. We usually
choose a value for $\dot M_*$ and solve equation~(\ref{SFR_const}) for
$t_{\rm burst}$. For an exponential SFR, we have
\begin{equation}
\dot M_*={M_*\over t_c}e^{-t/t_c}\,,
\label{SFR_expn}
\end{equation}

\noindent where $t_c=5\times10^7{\rm yrs}$ 
is the characteristic time. Since star formation never 
ends in this case, the stellar mass formed depends on the final time $t_f$. 
Equation~(\ref{SFR_expn}) is valid in the limit $t_f\gg t_c$.

There are some caveats about equations~(\ref{mbaryons}) and (\ref{mstar}).
The infalling gas must cool and form molecular clouds which then 
fragment into stars. If star formation is delayed until all the gas
has cooled, then equation~(\ref{mbaryons}) would be valid, but it
is more likely that star formation will start while a fraction of
the gas has not cooled yet. 
SNe and stellar winds from that first generation of stars
will inhibit the formation of subsequent generations
of stars, by reheating the ISM and possibly expelling some of it in the
form of a galactic outflow. Observations of galaxies
in the redshift range $0<z<4$ show that star formation
is a very inefficient process, with $M_*/M_{\rm gal}<0.03$ \citep{bcw10}.
Hence, the value of $M_b$ in equation~(\ref{mbaryons}) is an upper limit, 
which does not take into account the gas lost by galactic outflows.
As for the reheating of the gas and suppression of inflow, 
it is implicitly taken into
account in equation~(\ref{mstar}) by introducing a star formation efficiency
$f_*$. The mass of gas that was reheated and prevented from
forming stars is $M_{\rm reheat}=(1-f_*)M_b$.
In this paper, we consider star formation efficiencies $f_*=0.1$,
0.2, 0.5, and 1.0. Equations~(\ref{mbaryons}) and (\ref{mstar})
then give $M_*/M_{\rm gal}=0.016$, 0.032, 0.081 and 0.162, respectively.
The first two values are consistent with observations. The last
two values are quite extreme, and we considered them mostly to
investigate the behavior of the model under extreme conditions.

In this paper, we impose various
forms for the SFR to investigate the effect of the SFR on the
properties of the galactic outflows.
The effect of feedback and reheating of the ISM is all contained
implicitly in the value of $f_*$.
To provide a proper treatment of the aforementioned feedback processes,
we intend to modify the model such that the SFR
will be recalculated at every time step, from the physical conditions of
the ISM gas at that time, taking into account the effect of
all previous generation of stars. This will provide a consistent treatment
of feedback and self-regulating star formation, and eliminate the
need to specify a priori a star formation efficiency $f_*$.
This will be the subject of a forthcoming paper.

\subsection{Evolution of the Mass and Composition of the ISM}

Our algorithm tracks the evolution of $M_{\rm ISM_X}$, the mass of element
X contained in the ISM. 
At the beginning of each
timestep, the total mass of the ISM is given by
\begin{equation}
M_{\rm ISM}(t)=\sum_{\rm X} M_{\rm ISM_X}(t)\,,
\end{equation}

\noindent where the sum is over
all elements included in the algorithm, that is all elements from
hydrogen (${\rm X}=1$) to gallium (${\rm X}=31$). These quantities are
initialized at the beginning of the simulation, and updated during each
timestep $\Delta t$, as follows.
First, we calculate the mass ejected by stellar winds and SNe.
We include the contribution from all stars present at that time,
taking into account their {\it current} ages, 
initial metallicities, and initial masses:
\begin{eqnarray}
M_{\rm SW_X}(t)&=&\sum_k\dot M_{\rm SW_X}^{\rm SB99}(\tau_k,Z_k,M_k)\Delta t\,,
\label{mejectSW}
\\
M_{\rm SNe_X}(t)&=&\sum_k\dot M_{\rm SNe_X}^{\rm SB99}(\tau_k,Z_k,M_k)
\Delta t\,.
\label{mejectSN}
\end{eqnarray}

\noindent where $\tau_k$, $Z_k$, and $M_k$ are the current age, initial 
metallicity, and mass of population $k$, respectively.
The sums are over all the stellar populations that have
already formed by time $t$. The superscript SB99 indicates quantities 
calculated by Starburst99. We also calculate the total luminosity produced by 
SNe and stellar winds:
\begin{eqnarray}
L_{\rm SW}(t)&=&\sum_kL_{\rm SW}^{\rm SB99}(\tau_k,Z_k,M_k)\,,
\label{lumSW}
\\
L_{\rm SNe}(t)&=&\sum_kL_{\rm SNe}^{\rm SB99}(\tau_k,Z_k,M_k)\,.
\label{lumSN}
\end{eqnarray}

We then remove from the ISM the total mass of the
stars born during that timestep, and the mass removed by the galactic wind, 
and add the material ejected by SNe and stellar winds:
\begin{eqnarray}
&&M_{\rm ISM_X}(t+\Delta t)=M_{\rm ISM_X}(t)
-{M_{\rm ISM_X}(t)\over M_{\rm ISM}(t)}
\big[\dot M_*(t)\label{dmzdt}\nonumber\\
&&\qquad+\dot M_{\rm GW}(t)\big]\Delta t
+M_{\rm SW_X}(t)+M_{\rm SNe_X}(t)\,,
\end{eqnarray}

\noindent
where $\dot M_{\rm GW}(t)$
is the rate of mass loss by galactic wind (the calculation of
$\dot M_{\rm GW}$ is presented in the next section).
The mass remove from the
ISM by the star formation process
is used to generate several new stellar populations. Each population
is given an initial mass $M_k$, 
an initial metallicity $Z_k$ equal
to the metallicity $Z(t)$ of the ISM at that time, and we set the age $\tau_k$
of these new populations to zero. We then recompute the ISM metallicity:
\begin{eqnarray}
Z(t+\Delta t)&=&\Big[M_{\rm ISM}(t+\Delta t)-M_{\rm ISM_H}(t+\Delta t)
\nonumber\\
&&-M_{\rm ISM_{He}}(t+\Delta t)\Big]\Big/
M_{\rm ISM}(t+\Delta t)\,.
\end{eqnarray}

\noindent This expression accounts for the material removed from the ISM
by the star formation process and the galactic wind, 
and it also considers
the enriched material added by stars already formed. Finally, we
update the age of every stellar population:
\begin{equation}
\tau_k(t+\Delta t)=\tau_k(t)+\Delta t\,,\qquad\hbox{for all }k\,.
\end{equation}

\noindent These operations are repeated at every timestep in the simulation.
 
In our model, $M_{\rm ISM_X}$ is a function of time only.
This assumes that metals ejected
into the ISM by SNe and stellar winds are instantaneously mixed.
This commonly-used approximation has the advantage of being simple to 
implement.
However, in reality, it will take some finite time before the metals are
fully mixed. For this reason, the efficiency of metal-enrichment
of the IGM in our model should be considered as an upper limit.

Starburst99 cannot calculate directly the
mass loss and luminosities appearing in 
equations~(\ref{mejectSW}) to (\ref{lumSN}) for any initial metallicity $Z_i$.
It is limited to the values $Z_i=0.001$, 0.004, 0.008, and 0.02. 
We therefore need to interpolate the results of 
Starburst99 in order to get the quantities $\dot M^{\rm SB99}$
and $L^{\rm SB99}$. For values of $Z_i$ in the range $[0.001,0.02]$, we 
calculate $\dot M^{\rm SB99}$ and $L^{\rm SB99}$ at the metallicities 
immediately before and after $Z_i$, and interpolate between them,
using equations of the form
\begin{eqnarray}
\log\dot M^{\rm SB99}&=&A\log Z_i+B\,,
\label{interpol_dotM}
\\
\log L^{\rm SB99}&=&C\log Z_i+D\,.
\label{interpol_L}
\end{eqnarray}

\noindent
In the range $[0,0.001]$, we treat stellar winds and SNe differently.
We assume that, in that range, the mass loss
by stellar wind is proportional to $Z_i^\alpha$, with $\alpha\approx0.625$,
as in equations~(\ref{mej1}) to~(\ref{bmi}). Hence,
\begin{eqnarray}
\dot M_{\rm SW}^{\rm SB99}(Z_i)&=&\dot M_{\rm SW}^{\rm SB99}(0.001)
\left({Z_i\over0.001}\right)^{0.625}\,,\\
L_{\rm SW}^{\rm SB99}(Z_i)&=&L_{\rm SW}^{\rm SB99}(0.001)
\left({Z_i\over0.001}\right)^{0.625}\,.
\end{eqnarray}

\noindent
For SNe, we use the values $L_{\rm SNe}^{\rm SB99}(Z_i=0.001)$ 
at lower metallicities. This is a valid approximation because the
dependence of the stellar lifetimes
on metallicity is weak, as Figure~\ref{lum_SN} shows.

\subsection{Mass Loss by Galactic Wind}

The presence of a galactic wind enables a fraction of the ISM to escape
the galaxy and enrich the surrounding IGM. Since the wind is generated
by the thermal energy deposited in the ISM by stars, we expect the
mechanical energy of the galactic wind
to be proportional to the rate of energy
injection by SNe and stellar winds:
\begin{equation}
{1\over2}\dot M_{\rm GW}(t)V_{\rm GW}^2\propto L(t)\,,
\label{MproptoL0}
\end{equation}

\noindent where $\dot M_{\rm GW}$ and $V_{\rm GW}$ are the mass loss rate
by galactic wind and the velocity of the wind, respectively. The galactic wind
will create a cavity expanding into the IGM (see \S~4 below).
The expansion of the cavity is driven by the mechanical energy
$\dot M_{\rm GW}(t)V_{\rm GW}^2/2$ deposited into the IGM by the wind,
but does not depend separately on $\dot M_{\rm GW}$ and $V_{\rm GW}$.
Hence, to determine $\dot M_{\rm GW}$, we must make an additional assumption.
There are two limiting cases:
One limit consists of having $\dot M_{\rm GW}$ constant, in which
case increasing the number of SNe will increase the wind velocity.
The opposite limit consists of having
a constant wind velocity $V_{\rm GW}$, 
in which case an increase in the number of
SNe results in a larger amount of matter being ejected. 

It would take detailed high-resolution simulations to determine which
of these limits is correct. Ultimately,
the critical factor should be the spatial distribution of SNe. A single
SN will only affect the ISM located in its vicinity. In the case of several
SNe, there collective effect should critically depend on their level of 
clustering. If all SNe are concentrated in a same location, the same region
will be affected, and the net effect will be to eject the same matter, but
at a larger velocity. If instead the SNe are distributed throughout 
the galaxy, 
each SN will affect a different part of the ISM, and the net result will 
be to eject more material, but at the same velocity. This last case
is the limit we adopt
in this paper. Consequently, the reader should keep in mind that our 
estimates of the amount of material ejected is an upper limit. Under 
this assumption the rate of mass loss by the galactic wind is proportional
to the luminosity,
\begin{equation}
\dot M_{\rm GW}(t)\propto L(t)\,,
\label{MproptoL}
\end{equation}

To determine the constant of proportionality, we first integrate
the functions $\dot M_{\rm GW}(t)$ and $L(t)$ over the lifetime of the galaxy:
\begin{eqnarray}
M_{\rm GW}^{\rm tot}&=&\int_0^{t_f}\dot M_{\rm GW}\,dt\,,
\label{dotM_GW}\\
E^{\rm tot}&=&\int_0^{t_f}L\,dt\,,
\end{eqnarray}

\noindent where $M_{\rm GW}^{\rm tot}$ is the total mass
ejected into the galactic wind, and $E^{\rm tot}$ is the total energy
deposited in the ISM. We can rewrite equation~(\ref{MproptoL}) as
\begin{equation}
\dot M_{\rm GW}(t)={M_{\rm GW}^{\rm tot}\over E^{\rm tot}}L(t)\,.
\label{wind1}
\end{equation}

The problem is that $M_{\rm GW}^{\rm tot}$ and $E^{\rm tot}$ are not known
until the simulation is completed, and to perform the simulation, 
we need to know these quantities {\it in advance\/} in order to
calculate $\dot M_{\rm GW}$ at every timestep. To solve this problem,
we replace $M_{\rm GW}^{\rm tot}$ and $E^{\rm tot}$ in 
equation~(\ref{wind1}) by approximations that can be calculated
{\it ab initio}, before actually performing the simulation.

\subsubsection{Estimate of $E^{\rm tot}$.}

\begin{figure}
\begin{center}
\includegraphics[width=3.3in]{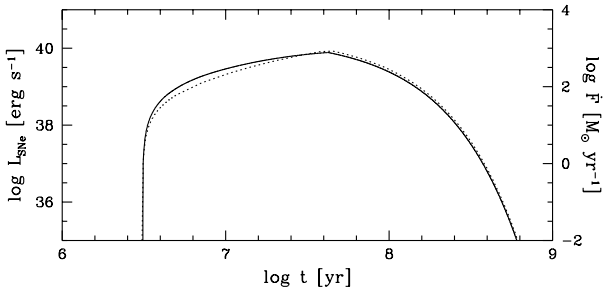}
\caption{Solid curve: SNe luminosity versus time, for a $10^9\msun$
galaxy with an exponential SFR and a star formation efficiency $f_*=0.1$.
Dotted curve: Integrated mass loss rate $\dot F(t)$, for the same galaxy.}
\label{lum_t}
\end{center}
\end{figure}

\noindent The luminosity $L(t)$ is calculated as the simulation proceeds,
but we can estimate it as follows:
first, as we shall see below, the contribution of stellar winds
to the luminosity becomes negligible once the SNe phase starts. 
If we neglect stellar winds, and also neglect the weak dependence of the SN
luminosity on the metallicity (see Fig.~\ref{lum_SN}), we can
directly estimate the luminosity from the star formation rate $\dot M_*$. We
define an {\it integrated mass loss rate} $\dot F(t)$ using:
\begin{equation}
\dot F(t+t_{\rm onset})\equiv\int_{t-t_{\rm active}}^t\dot M_*(t')dt'\,,
\end{equation}

\noindent where $t_{\rm onset}$ is the time elapsed between the formation
of the stellar population and the onset of the first SN, and
$t_{\rm active}$ is the time duration of the SNe phase.\footnote{The lifetimes
of the shortest-lived and longest-lived progenitors are therefore
$t_{\rm onset}$ and $t_{\rm onset}+t_{\rm active}$, respectively.}
Figure~\ref{lum_t} shows the luminosity $L(t)$ obtain from the simulation,
and the quantity $\dot F(t)$ calculated using
$t_{\rm onset}=3.11\,{\rm Myr}$ and $t_{\rm active}=39.2\,{\rm Myr}$
(Fig.~\ref{lum_SN}).
To a very good approximation, $\dot F(t)$ is proportional to $L(t)$.
We can therefore approximate equation~(\ref{wind1}) as
\begin{equation}
\dot M_{\rm GW}={M_{\rm GW}^{\rm tot}\over F^{\rm tot}}\dot F(t)\,,
\label{wind2}
\end{equation}

\noindent where
\begin{equation}
F^{\rm tot}=\int_0^{t_f}\dot F\,dt\,.
\end{equation}

\noindent Both $\dot F(t)$ and $F^{\rm tot}$ are calculated at the beginning
of the simulation. 

\subsubsection{Estimate of $M_{\rm GW}^{\rm tot}$.}

We still need to determine the total mass ejected by the galactic wind,
$M_{\rm GW}^{\rm tot}$, to be able to use equation~(\ref{wind2}). Like
the total energy deposited in the ISM,
$E^{\rm tot}$, $M_{\rm GW}^{\rm tot}$ is not known until the simulation
is completed. To estimate it, we
replace $\dot M_{\rm GW}$ in equation~(\ref{dotM_GW})
by an approximation that can be calculated at the beginning of the simulation.
We then integrate to get $M_{\rm GW}^{\rm tot}$, we substitute that value
in equation~(\ref{wind2}), which then provides
the mass loss by galactic wind during the simulation. To find an initial
approximation for $\dot M_{\rm GW}$, we first notice that observations 
at different redshifts suggest a relation between the mass loss by galactic 
winds and the star formation rate \citep{martin99}, often expressed
 in terms of the ratio
\begin{equation}
\eta\equiv{\dot M_{\rm GW}\over\dot M_*}\,.
\end{equation}

\noindent This value appears to vary significantly among galaxies,
with values ranging from 0.01 to 10 \citep{vcbh05}. \citet{mqt05} derived
analytical relations between the factor $\eta$ and the velocity dispersion
$\sigma$, for both momentum-driven and energy-driven winds. We focus in 
this paper on energy-driven winds, but will consider momentum-driven winds in
future work. For energy-driven winds, \citet{mqt05} derived the
following relation:
\begin{equation}
\dot M_{\rm GW}=\dot M_*\,\xi_{0.1}\,\varepsilon_3^{\phantom1}
\left({300\,{\rm km\,s^{-1}}\over\sigma}\right)^2\,,
\label{dotMGW_murray1}
\end{equation}

\noindent where $\sigma$ is the velocity dispersion,
$\varepsilon_3^{\phantom1}\equiv1000E^{\rm tot}/M_*c^2$, 
and $\xi_{0.1}\equiv f_w/0.1$, with $f_w(M_{\rm gal})$ the fraction of 
energy provided by stars that is used to power the wind, 
for a galaxy of mass $M_{\rm gal}$ \citep{sfm02}.
To calculate $\varepsilon_3^{\phantom1}$, 
we use Starburst99 with a $10^6\msun$ 
stellar population and a standard IMF.
The total energy $E^{\rm tot}$ produced by SNe and 
stellar winds is always of the order of $10^{55.3}{\rm ergs}$, for all
metallicities. This gives $\varepsilon_3^{\phantom1}=0.011$. 
Equation~(\ref{dotMGW_murray1}) reduces to
\begin{equation}
\dot M_{\rm GW}=0.11\dot M_*f_w
\left({300\,{\rm km\,s^{-1}}\over\sigma}\right)^2\,.
\label{dotMGW_murray2}
\end{equation}

\noindent For a galaxy of mass $M_{\rm gal}$, the velocity dispersion
$\sigma$ is calculated using the equation of \citet{od08}:
\begin{equation}
\sigma=200\left[{M_{\rm gal}\over5\times10^{12}\msun}h
{H(z_{\rm gf})\over H_0}\right]^{1/3}{\rm km\,s^{-1}}\,,
\label{sigma}
\end{equation}

\noindent where $z_{\rm gf}$ is the formation redshift of the
galaxy, $H$ is the Hubble parameter, with $H_0$ being its present value,
and $h=H_0/100\,\rm km\,s^{-1}\,Mpc^{-1}$.
Equations~(\ref{dotMGW_murray2}) and~(\ref{sigma})
give us an estimate of $\dot M_{\rm GW}$. We then apply
equation~(\ref{dotM_GW})
to calculate $M_{\rm GW}^{\rm tot}$, which we substitute
in equation~(\ref{wind2}). This equation is then used to calculate the
mass loss by galactic wind during the simulation [eq.~(\ref{dmzdt})].

\begin{figure}
\begin{center}
\includegraphics[width=3.3in]{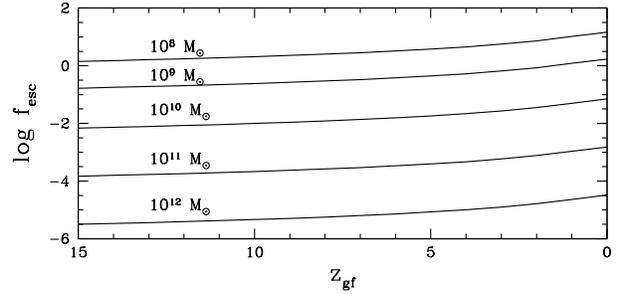}
\caption{Escape fraction versus redshift, for a $10^9\msun$
galaxy with a star formation efficiency $f_*=0.1$.}
\label{fesc1}
\end{center}
\end{figure}

The parameter $f_{\rm esc}$ is defined as the fraction of the ISM mass that
escapes the galaxy:
\begin{equation}
f_{\rm esc}={M_{\rm GW}^{\rm tot}\over M_b}={\Omega_0\over\Omega_{b0}}
{M_{\rm GW}^{\rm tot}\over M_{\rm gal}}\,.
\label{fesc}
\end{equation}

\noindent 
Figure~\ref{fesc1} shows
$f_{\rm esc}$ versus formation redshift $z_{\rm gf}$, for various
galactic masses $M_{\rm gal}$, with a star 
formation efficiency of $f_*=0.1$. The dependence on $z_{\rm gf}$ comes 
entirely from the factor $H(z_{\rm gf})$ in equation~(\ref{sigma}). Galaxies
that form earlier have a larger velocity dispersion $\sigma$ for a
given mass $M_{\rm gal}$. Lower-mass galaxies eject a larger fraction of
their ISM than higher mass galaxies, and for a $10^8\msun$ galaxy, 
we get $f_{\rm esc}>1$, which simply means that
the entire ISM will be ejected from the galaxy (so the actual $f_{\rm esc}$
is unity).

\section{GALACTIC WIND MODEL}

\subsection{The Dynamics of the Expansion}

Tegmark et al. (1993, hereafter TSE) presented
a formulation of the expansion of isotropic galactic winds
in an expanding universe.
In this formulation, the injection of thermal energy
produces an outflow of radius $R$, which consists of a dense shell
of thickness $R\delta$ containing a cavity.
A fraction $1-f_m$ of the mass of the gas is piled up in the shell, while a
fraction $f_m$ of the gas is distributed inside the cavity.
We normally assume $\delta\ll1$, $f_m\ll1$, that is, 
most of the gas is located inside a thin shell.
This is called the {\it thin-shell approximation}.

The evolution of the shell radius $R$ expanding out of a halo
of mass $M_{\rm gal}$, is described
by the following system of equations:
\begin{eqnarray}
\label{rdotdot}
\ddot R&=&{8\pi G(p-p_{\rm ext})\over\Omega_bH^2R}-{3\over R}(\dot R-HR)^2
-{\Omega H^2R\over2}\nonumber\\
&&-{GM_{\rm gal}\over R^2}\,,\\\label{pdot}
\dot p&=&{L\over2\pi R^3}-{5\dot Rp\over R}\,,
\end{eqnarray}

\noindent where a dot represents a time derivative, $\Omega$, $\Omega_b$,
and $H$ are the total density parameter, baryon density parameter, and
Hubble parameter at time $t$, respectively,
$L$ is the luminosity, $p$ is the pressure inside the cavity
resulting from this luminosity, and $p_{\rm ext}$ is the external
pressure of the IGM. 
The four terms in equation~(\ref{rdotdot}) represent, from left to right, the 
driving pressure of the outflow, the drag due to sweeping up the IGM and 
accelerating it from velocity $HR$ to velocity $\dot R$, and the gravitational
deceleration caused by the expanding shell and by the halo itself.
The two terms in equation~(\ref{pdot}) represent the increase
in pressure caused by injection of thermal energy, and the drop in 
pressure caused by the expansion of the wind, respectively.

The external pressure, $p_{\rm ext}$, depends upon the density and
temperature of the IGM. As in PMG07, we will 
assume a photoheated IGM made of ionized hydrogen
and singly-ionized helium (mean molecular mass
$\mu=0.611$), with a fixed temperature
$T_{\rm IGM}=10^{4}{\rm K}$ \citep{mfr01} and 
an IGM density equal to the mean baryon density
$\bar\rho_b$. The external pressure at redshift $z$ is then given by:
\begin{equation}
\label{pext}
p_{\rm ext}(z)={\bar\rho_bkT_{\rm IGM}\over\mu}=
{3\Omega_{b,0}H_0^2kT_{\rm IGM}(1+z)^3\over8\pi G\mu}\,.
\end{equation}

The luminosity $L$ is the rate of energy deposition or 
dissipation within the wind and is given by:
\begin{equation}
L(t)=f_w(L_{\rm SNe}+L_{\rm SW})-L_{\rm comp}\,,
\end{equation}

\noindent where $L_{\rm SNe}$ and $L_{\rm SW}$ are the total luminosity
responsible for generating the wind, as given by 
equations~(\ref{lumSN}) and (\ref{lumSW}), respectively.
$L_{\rm comp}$ represents
the cooling due to Compton drag against CMB photons and is given by:
\begin{equation}
\label{lcomp2}
L_{\rm comp}={2\pi^3\over45}{\sigma_t\hbar\over m_e}
\left({kT_{\gamma0}\over\hbar c}\right)^4(1+z)^4pR^3\,,
\end{equation}

\noindent 
where $\sigma_t$
is the Thomson cross section, and $T_{\gamma0}$ 
is the present CMB temperature.

The expansion of the wind is initially driven by the luminosity.
After the SNe turn off, the outflow enters the 
``post-SN phase.''\footnote{The stellar winds are still on, but 
their contribution is negligible at this point.}
The pressure inside the wind keeps driving the expansion, but this
pressure drops since there is no energy input from SNe.
Eventually, the pressure will drop down to the level of the external IGM 
pressure. At that point, the expansion of the wind will simply follow the
Hubble flow. 

\subsection{Metal Distribution inside the Galactic Wind}

In the TSE model, the baryon density inside the cavity
is $\rho_i=\rho_b(t)f_m/(1-\delta)^3$, while
the baryon density inside the shell is 
$\rho_s=\rho_b(t)(1-f_m)/[1-(1-\delta)^3]$. This gives a mass 
$M=4\pi R^3\rho_b(t)/3$ inside the volume of
radius $R$, which is precisely the mass of
the IGM within that radius in the absence of a wind. Therefore, in the
TSE model, the material inside the shell is swept IGM material, while the
material inside the cavity is IGM material left behind. {\it The mass $M_{GW}$
added by the galactic wind is neglected in the TSE model.}
Hence, the TSE model does not predict the distribution of that mass inside
the cavity. This means that any distribution we chose would not violate
the assumptions on which the TSE model is based.

The simplest approximation for the distribution of metals in the wind
consists of assuming that the metals carried 
by the galactic wind are spread evenly inside the cavity (see
\citealt{sfm02}; PMG07; \citealt{bmg11}). 
This poses a problem for the metals ejected near
the end of the post-SN phase, just before the wind joins
the Hubble flow. These metals would have to be carried across the
entire radius of the cavity, at velocities that exceed the wind velocity.
Processes such as turbulence and diffusion could homogenize the distribution
of metals inside the cavity, but only over a finite time period.
In this paper, we take the opposite approach, by assuming no mixing. Hence,
the gas that escapes the galaxy early on will travel larger distances
than the gas that escapes later. Since the metallicity and composition of 
the ISM evolves with time, the galactic wind will acquire both a
metallicity gradient and a composition gradient, with the inner parts
containing a larger proportion of metals.
To simulate such wind, we use a system of concentric spherical shells.
At the end of
every timestep, the code calculates the amount of gas that will be
added to the galactic wind:
\begin{equation}
\Delta M_{\rm GW}(t_i)=\dot M_{\rm GW}(t_i)\Delta t\,,
\end{equation}

\noindent where $t_i$ is the time corresponding to the timestep. 
After the first time step, the wind reaches a radius 
$R_1\equiv R_{\rm GW}(\Delta t)$.
We deposit the wind material produced during that time step into the sphere
of radius $R_{\rm GW}(\Delta t)$, which constitutes our
central shell. After the second timestep, the wind
now reaches radius $R_2\equiv R_{\rm GW}(2\Delta t)$. We first transfer 
the wind material located between 0 and $R_1$ into a shell of
inner radius $R_1$ and outer radius $R_2$, and we then deposit 
the wind material produced during the second timestep into the central shell.
This process is then repeated. At every timestep $n$, a new shell is created
between radii $R_n$ and $R_{n-1}$, all the wind material is shifted outward
by one shell, and the new material is deposited into the central shell.
Finally, when the wind enters the post-SN phase, we no longer deposit
materiel into the wind, and the shells expand homologously with the 
cavity. One nice feature of this model is that it
has no free parameter. In particular, if does not depend on the 
value of the timestep. Using a different timestep would change 
the resolution at which the wind profile is determined, 
but not the profile itself.

\section{RESULTS}

Here we use the algorithm described in \S4 to study the evolution of
starburst galaxies, in a concordance $\Lambda$CDM universe 
with density parameter $\Omega_0=0.27$, 
baryon density parameter $\Omega_{b0}=0.044$, cosmological constant
$\lambda_0=0.73$, and Hubble constant $H_0=71\,{\rm km\,s^{-1}Mpc^{-1}}$
($h=0.71$).
Because the parameter space is large, we focus on
a fiducial case: a dwarf galaxy of mass $M_{\rm gal}=10^9\msun$ forming
at $z_{\rm gf}=15$.
This case is particularly important because the vast
majority of galaxies in the universe are dwarfs, and in CDM cosmology, these
galaxies tend to form at high redshift (e.g. \citealt{bfpr84}). 
Also our model assumes that 
galaxies form by monolithic collapse and not by the merger of well-formed
galaxies, an assumption that is more appropriate for dwarfs
(e.g. \citealt{bfpr84}).
Note that the value of $z_{\rm gf}$ matters in the model. It affects 
the expansion of the galactic wind, which in
turns affects the evolution of the ISM. 

\begin{figure}
\begin{center}
\includegraphics[width=3.3in]{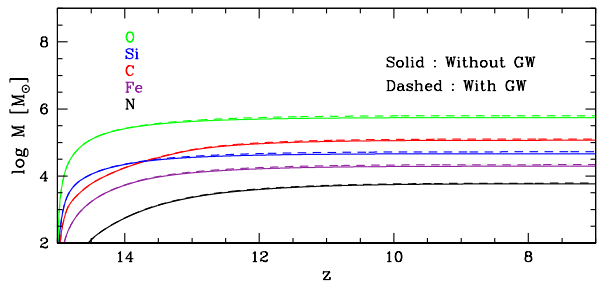}
\caption{Mass of some elements present in the ISM
versus redshift, for a $10^9\msun$
galaxy with a star formation efficiency $f_*=0.1$ and
an exponential SFR. The colors corresponds to various elements, 
as indicated.
Solid lines: simulation without galactic wind;
dashed lines: simulation with galactic wind.}
\label{massZ_z}
\end{center}
\end{figure}

We performed two simulations, one with our basic model, and another one in
which we turned off the galactic wind.
Figure~\ref{massZ_z} shows the abundances of a few 
elements in the ISM. 
Most of the ISM enrichment occurs between redshifts $z=15$ and $z=13$,
during the epoch of intense SNe activity. At lower redshifts, the enrichment
by stellar winds dominates.
The effect of the galactic wind is very small. Adding the wind results in
a 10\% increase in ISM metallicity, caused by the removal of low-metallicity
ISM during the early stages of the wind.
The effect the galactic wind can be much more significant
but this requires a smaller galactic mass $M_{\rm gal}$ or a larger
star formation efficiency $f_*$, or SFR much more extended in time than 
the ones we have considered.

In the next three subsections, we explore the parameter space by varying,
respectively, the SFR, the star formation efficiency $f_*$, and
the mass $M_{\rm gal}$ of the galaxy. 

\subsection{Star Formation Rate}

\begin{figure}
\begin{center}
\includegraphics[width=3.3in]{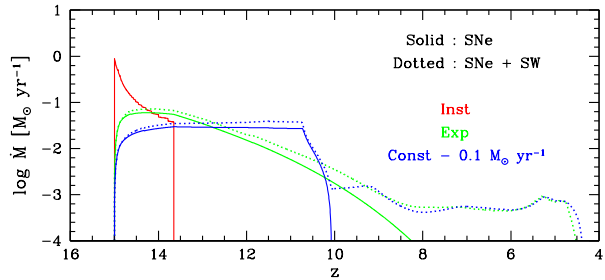}
\caption{Mass loss rate of stars versus redshift, for a $10^9\msun$
galaxy with a star formation efficiency $f_*=0.1$. 
The various colors represent different SFRs, as indicated. 
Solid lines: simulations with SNe only; dotted lines: 
simulations with SNe and stellar winds.}
\label{massloss}
\end{center}
\end{figure}

Figure~\ref{massloss} 
shows the mass returned to the ISM by stellar winds and SNe,
versus redshift, for the different SFRs. With an instantaneous SFR,
there is only one stellar population and the mass loss profile is 
identical to the one provided by SB99. Stellar winds are absent in this 
case because all the stars were formed in a metal-free ISM.
For the constant SFR, the increase in mass returned to the ISM is caused by
the formation of more and more stars. Since SNe dominate over stellar 
winds, a plateau is eventually reached when the time of the simulation is
equal to the lifetime of the SNe for the first generation of stars.
After that moment, the contribution of a new population is compensated
by the death 
of an old population. The processus is the same for the exponential
SFR, except that the mass of the stellar populations decreases with time. 
Hence, the death of an old population is replaced by the birth of
a less-massive
population, which explains the absence of a plateau. For the constant SFR,
there is a sudden drop at $z=10.7$ which corresponds to the last SNe
explosions. The material ejected after that 
corresponds to the giant phase of low-mass stars.

\begin{figure}
\begin{center}
\includegraphics[width=3.3in]{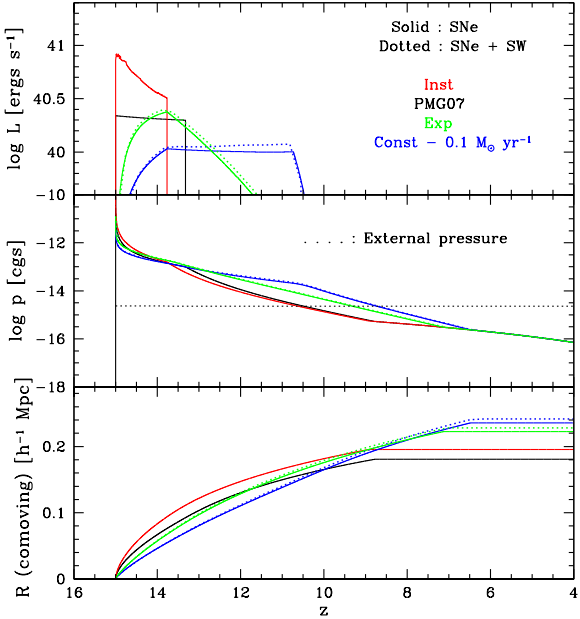}
\caption{Luminosity (top panel), internal pressure (middle panel), 
and comoving radius of the galactic wind (bottom panel) versus
redshift, for a $10^9\msun$
galaxy with a star formation efficiency $f_*=0.1$. Colors and linetypes
have the same meaning as in Figure~\ref{massloss}. The 
solid black lines show,
for comparison, the model of PMG07. The dotted black line in the
middle panel shows the external pressure of the IGM.}
\label{pressure}
\end{center}
\end{figure}

\begin{figure}
\begin{center}
\includegraphics[width=3.3in]{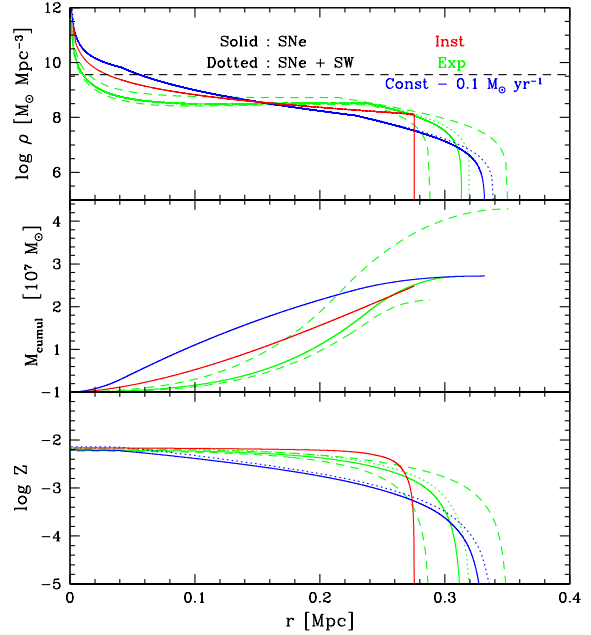}
\caption{Density profile (top), cumulative mass profile (middle),
and metallicity profile (bottom) of material
ejected into the IGM by the galactic wind, at $z=0$,
for a $10^9\msun$
galaxy with a star formation efficiency $f_*=0.1$. Colors and linetypes
have the same meaning as in Figures~\ref{massloss}
and~\ref{pressure}. The dashed curves show the results
for an exponential SFR with minimum SNe progenitor masses
$M_i=6\msun$ and $M_i=10\msun$.
In the top panel, the horizontal dashed line shows the density of the IGM
inside the cavity, assuming $f_m=0.1$.}
\label{eject_MIG}
\end{center}
\end{figure}

Figure~\ref{pressure} shows the luminosity, internal pressure, and 
comoving radius of
the galactic wind, for the various SFRs. Again, stellar winds do not have 
much effect on the results.
One interesting aspect is that an extended period of star formation tends to
produce a larger final radius for the outflow, compared with an instantaneous
SFR, even though the total stellar mass $M_*$ formed is the same.
The top panel of Figure~\ref{eject_MIG} shows the density profiles 
of the galactic wind. The density
gradients are very strong, with the density dropping by $3-5$ orders of
magnitude from the center to the edge. This is caused
mostly by the dilution resulting from the expansion.
The outer density profile is lower for the constant SFR than
for the instantaneous and exponential ones.
In our galactic wind model, the outer parts of the
wind contain gas that was expelled by the galaxy at early time.
The amount of material ejected during the early phases will
depend of the mass loss rate 
$\dot M_{\rm GW}$ at that time, which is proportional
to $L(t)$ [eq.~(\ref{wind1})]. As Figure~\ref{pressure} shows, in
the early phases, $L(t)$ is larger for the instantaneous and
exponential SFR's than for the constant SFR, which leads
to a larger amount of material being ejected, material which ends up in
the outer parts of the wind.
The dashed line shows the density of the IGM inside the cavity, assuming 
$f_m=0.1$. Not surprisingly, the wind density exceeds the IGM density inside
the galaxy, or immediately outside it. But at larger radii, the wind
density drops several orders of magnitude below the IGM density. 
We calculated the mass of the IGM inside the cavity, assuming a shell
thickness $\delta=0.05$. For the 5 cases plotted in Figure~\ref{eject_MIG}, 
the values are in the range $3.86-7.27\times10^8\msun$. The mass added by
the galactic wind is $2.73\times10^7\msun$, or between 3.8\% and 7.1\% of the
mass in the cavity. This justifies {\it a posteriori\/} the assumption made by 
TSE that the mass added by the wind can be neglected.

\begin{figure*}
\begin{center}
\includegraphics[width=6.0in]{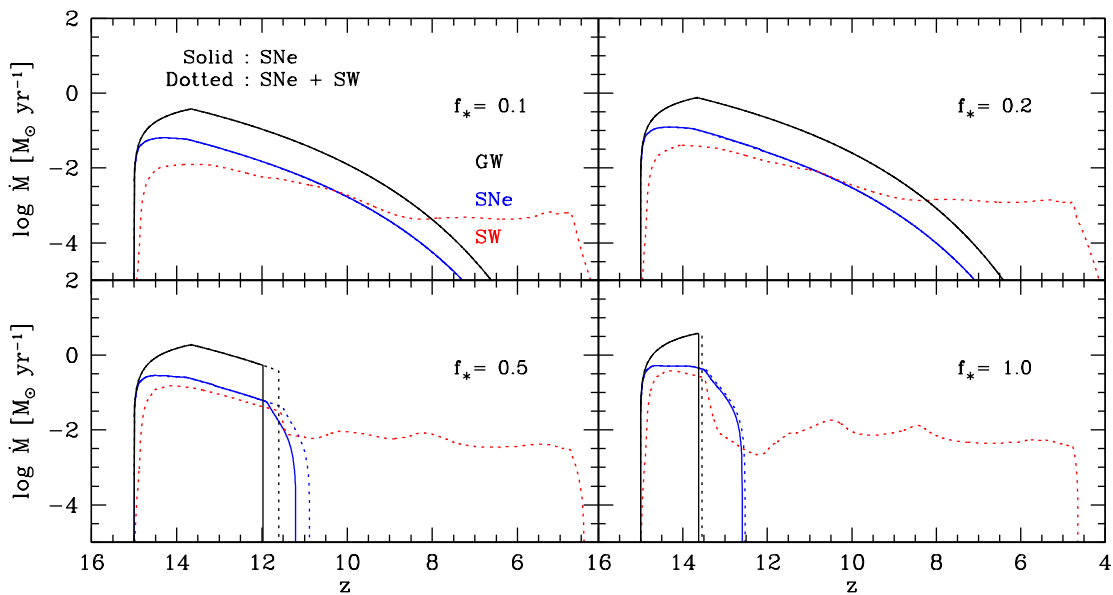}
\caption{Rate of mass loss by SNe (blue), stellar winds (red),
and galactic wind (black) versus redshift, for a $10^9\msun$
galaxy with an exponential SFR. 
Solid lines: simulations with SNe only;
dotted lines: simulations with SNe and stellar winds.
The various panels correspond to different
star formation efficiencies $f_*$, as indicated.
In the top panels, the solid and dotted lines are indistinct
for the galactic winds and SNe.}
\label{massloss2}
\end{center}
\end{figure*}

The middle panel of Figure~\ref{eject_MIG} shows the cumulative mass
profile of the galactic wind, that is, the mass $M_{\rm GW}(r)$
between 0 and $r$.
Even though the density is maximum in the center, the actual amount
of ejecta located near the galaxy is negligible. For a galaxy of
mass $10^9M_\odot$, collapsing at redshift $z=15$, the virial radius is
$r_{200}=2\,{\rm kpc}$, and the radius of the stellar component is even
smaller. Essentially all the gas contained in the galactic wind has 
been ejected from the galaxy, and most of it is located at radius
$r>100\,{\rm kpc}$.

The bottom panel of Figure~\ref{eject_MIG} shows the metallicity profiles.
The metallicity gradients have a different
origin, since the dilution caused by the expansion of the wind equally 
affects metals, hydrogen, and helium.
Since the material located in the outer parts of the wind was ejected
earlier than material located in the inner parts, the metallicity gradient
simply reflects the time-evolving chemical composition of the ISM. 
With an instantaneous SFR, the ISM is enriched in metals very rapidly. Hence,
the gas ejected into the galactic wind at early times is already metal-rich.
As a result, the metallicity in the outer parts of the wind is
larger for the instantaneous SFR than for the other SFRs.

The dashed curves in Figure~\ref{eject_MIG} show the effect of changing
the minimum SNe progenitor mass (for an exponential SFR). 
Lowering the minimum mass from $8\msun$ to $6\msun$ increase the final radius 
of the outflow by 12\% and the total mass ejected by 59\%.
Increasing the minimum mass to $10\msun$ reduces the final radius of
the outflow by 8\% and the total mass ejected by 21\%.

\subsection{Efficiency of the Star Formation}

\begin{figure}
\begin{center}
\includegraphics[width=3.3in]{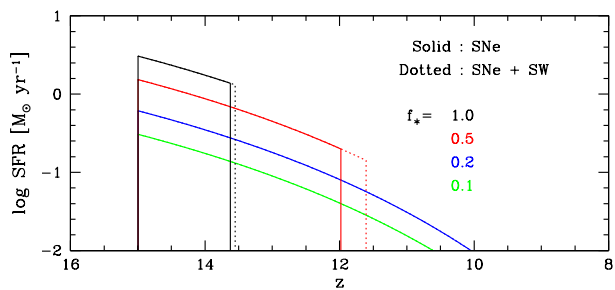}
\caption{Rate of mass converted to stars
versus redshift, for a $10^9\msun$
galaxy with an exponential SFR. 
Solid lines: simulations with SNe only;
dotted lines: simulations with SNe and stellar winds.
The various colors 
correspond to different star formation efficiencies $f_*$, as indicated.
For $f_*=0.2$ and 0.1, the solid and dotted lines are indistinct.}
\label{mass2star}
\end{center}
\end{figure}

Figure~\ref{massloss2}
shows the effect of varying the efficiency of star 
formation for a $10^9\msun$ galaxy with an exponential SFR. Apart
from the fact that the mass loss rate increases with the number of stars 
formed, the most striking feature of this figure is that,
for $f_*\geq0.5$, the galactic wind
can be sufficiently powerful to eject the totality of the ISM.
Eventually, the ISM is replenished by SN
ejecta and stellar winds produced by stars already formed. 
Figure~\ref{mass2star} shows that
stellar winds are more significant
with $f_*=0.5$ than with $f_*=1$. Lowering $f_*$ spreads
star formation over a longer period of time, enabling stars to
form in an environment richer in metals, and resulting in stronger 
stellar winds.

\begin{figure}
\begin{center}
\includegraphics[width=3.3in]{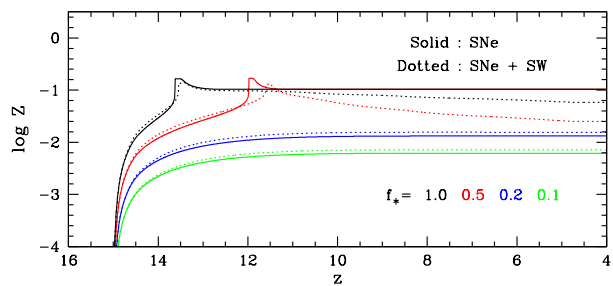}
\caption{Metallicity of the ISM versus redshift, for a $10^9\msun$
galaxy with an exponential SFR. 
Colors and linetypes have the same meaning as in Figure~\ref{mass2star}.}
\label{ZISM_z}
\end{center}
\end{figure}

Figure~\ref{ZISM_z} shows the evolution of the ISM metallicity.
The metallicity increases faster with a higher star formation efficiency,
since there are more stars available to enrich the ISM. When all the gas in
the galaxy is eventually ejected into the IGM, the metallicity
experiences a sudden increase before reaching a maximum value. This sudden
increase occurs when the mass remaining into the ISM becomes similar to
the mass returned by stars. Then, the mass of the ISM keeps dropping, and 
the metallicity approaches the value corresponding to the last stellar 
ejecta. Afterward, the metallicity decreases because the last SNe
ejected fewer and fewer metals.
Then, when stellar winds
are taken into account, the metallicity of the ISM continue to decrease
with time because low-mass stars in their giant phases eject material composed
mostly of hydrogen and helium.

\begin{figure*}
\begin{center}
\includegraphics[width=6.0in]{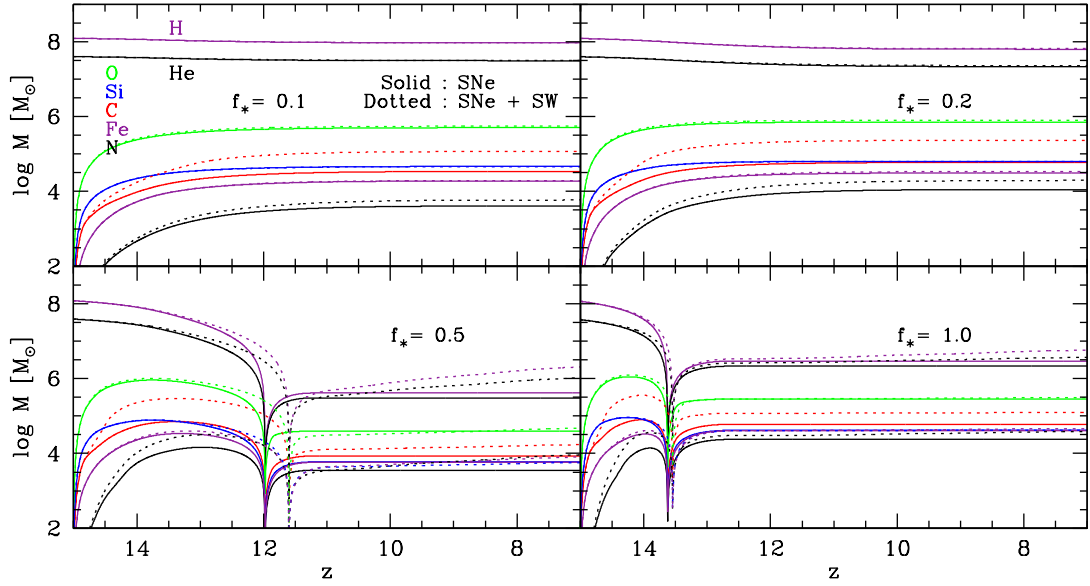}
\caption{Mass of various elements present in the ISM versus redshift, 
for a $10^9\msun$ galaxy with an exponential SFR.
The various panels correspond to different
star formation efficiencies $f_*$, as indicated.
The colors corresponds to various elements, 
as indicated.
Solid lines: simulations with SNe only;
dotted lines: simulations with SNe and stellar winds.}
\label{metals_z}
\end{center}
\end{figure*}

\begin{figure*}
\begin{center}
\includegraphics[width=6.0in]{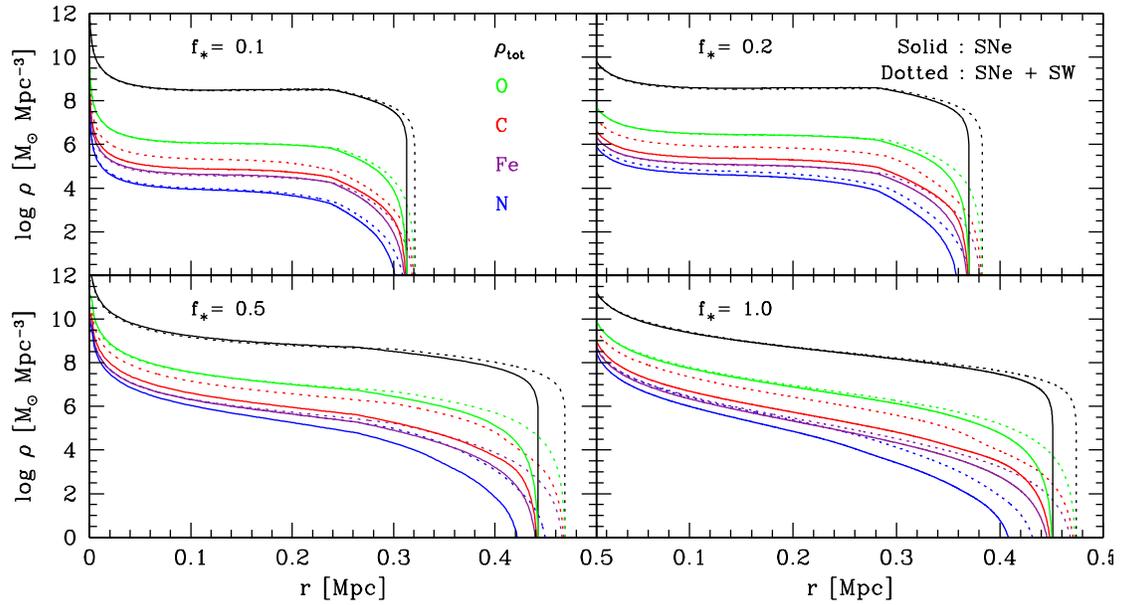}
\caption{Density profile of material ejected in the IGM,
for a $10^9\msun$ galaxy with an exponential SFR.
Black lines: total density; colored lines: density of various elements,
as indicated.
Linetypes have the same meaning as in Figure~\ref{metals_z}.}
\label{M_IGM}
\end{center}
\end{figure*}

\begin{figure}
\begin{center}
\includegraphics[width=3.3in]{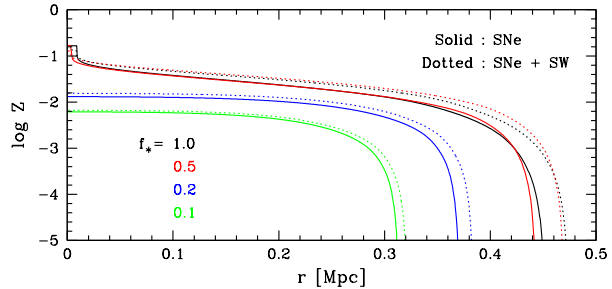}
\caption{Metallicity profile of material ejected in the IGM at $z=0$,
for a $10^9\msun$
galaxy with an exponential SFR. The various colors
represent different star formation efficiencies $f_*$. 
Colors and linetypes have the same meaning as in 
Figures~\ref{mass2star} and~\ref{ZISM_z}.
}
\label{Z_IGM}
\end{center}
\end{figure}

Figure~\ref{metals_z} shows the evolution of the composition
of the ISM. The importance of stellar winds becomes naturally larger
when the star formation efficiency increases. 
The stellar winds do not have much effect on the
total mass of the ISM. However, there is a significant difference in the
abundances of carbon and nitrogen for $f_*>0.1$. 
Figure~\ref{M_IGM} shows the density of various elements inside the
galactic wind. The external part of the
galactic wind has the composition of the ISM during the early phases.
To have a significant 
effect on the composition of the external regions of the galactic wind, 
which are the prime contributor to the IGM enrichment,
the enrichment of the ISM
must happen rapidly, which is the case when $f_*$ is large.
Figure~\ref{Z_IGM} shows th metallicity profile of the galactic wind,
for the various values of $f_*$.
The amount of metal ejected into the IGM seems large enough 
to fit observations.
Various studies indicate that the IGM metallicity
at redshifts between 2.5 and 3.5 is
of the order of $10^{-2.5}Z_\odot$ 
\citep{sc96,hellstenetal97,rauchetal97,daveetal98}. 

\subsection{The Mass of the Galaxy}

Usually, the evolution of the ISM is unaffected by the mass of the host galaxy.
Increasing the mass of the galaxy increases the mass of the ISM, the
mass in stars, the amount of gas ejected by SNe and stellar winds, and the 
amount of metals ejected by exactly the same factor.
The only thing that might affect this tendency is the mass loss
caused by the galactic wind. If the power of the wind is moderate,
the evolution of the ISM will be unaffected. This is the case for the most
massive galaxies, because the energy deposited is less and less coherent,
which reduces the 
fraction of energy $f_w$ used to produce the galactic wind. 
For less massive galaxies, the ISM is more enriched,
because the galactic wind expels more gas, which increases the relative
importance of metals returned by stars.

\begin{figure}
\begin{center}
\includegraphics[width=3.3in]{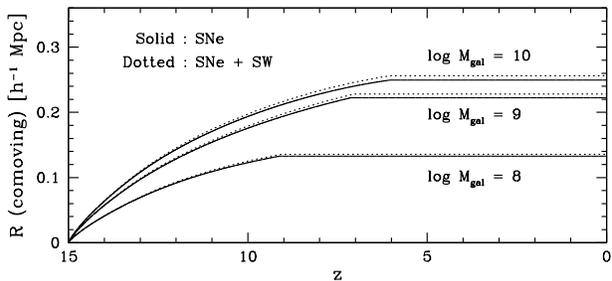}
\caption{Comoving radius of galactic wind versus redshift, for
galaxies with an exponential SFR and a star formation efficiency $f_*=0.1$,
for various galaxy masses $M_{\rm gal}$ in solar masses. 
Solid lines: simulations with SNe only; dotted lines: 
simulations with SNe and stellar winds.}
\label{R_mgal}
\end{center}
\end{figure}

Figure~\ref{R_mgal} shows the comoving radius of the galactic wind,
for galaxies of various masses, all  
having $f_*=0.1$ and an exponential SFR.
The final radius increases with the mass, but this effect is weak
and gets weaker at larger masses. The radius of the
galactic wind $R$ increases by a factor of 1.6 from
$10^8\msun$ to $10^9\msun$, and 1.13 from $10^9\msun$ to $10^{10}\msun$.
This results from the competition between several effects. The 
energy deposited into the ISM by SNe and stellar winds increases
linearly with $M_{\rm gal}$, but the fraction $f_w$ of that energy which
is used to power the wind decreases with $M_{\rm gal}$. 
At large masses, $f_w\propto1/M_{\rm gal}$ and the two effects cancel out.
At smaller masses, $f_w$ decreases slower than $1/M_{\rm gal}$ and
the energy available to power the wind increases with mass. That energy
must compete with the gravitational pull of the galaxy [last term in
eq.~(\ref{rdotdot})], which increases with galactic mass. This effect reduces
further the final radius of the wind, and at large masses $R$ actually
decreases with increasing mass. We do not include masses
$M_{\rm gal}=10^{11}\msun$ and $10^{12}\msun$ in
Figure~\ref{R_mgal}, because the galactic wind does not even start
for those objects.
With an instantaneous SFR, we can maximize the effect of the energy deposition
and produce a wind from $M_{\rm gal}=10^{11}\msun$, but 
this wind remains gravitationally bound to the 
galaxy and eventually falls back. 

\section{SUMMARY AND CONCLUSION}

We have combined a population synthesis code, interpolation tables
for the mass and composition of SN ejecta, and an analytical model
for galactic winds into a single algorithm that self-consistently
describes the evolution of starburst galaxies. This model describes 
the evolution of the stellar populations in the galaxy, the
evolution of the mass and chemical composition of the ISM, 
the propagation of the galactic
wind, and the distribution and abundances of metals inside the galactic 
wind. In particular, the algorithm (1) provides a detailed calculation of 
the energy deposited into the ISM by SNe and stellar winds, which is 
responsible for driving the galactic wind, 
(2) takes into account the time-evolution
of the chemical composition of the ISM, which directly affect the 
composition of the galactic wind, and (3) takes into account the removal of
the ISM by galactic winds, which affects the metallicity of
the ISM, and the metallicity of the stellar populations to follow.

Our first results concern the SFR for the galaxy. For a given
star formation efficiency $f_*$, a longer SFR tends to produce a galactic
wind that reaches a larger extent, but this wind will be less dense. 
By increasing the star formation efficiency,
we can produce a wind that reaches a larger extent and
has a higher metallicity near its front. 
In some cases, the energy deposited
by the stars is sufficient to completely expel the ISM. When it happens,
star formation is shut down, and the galactic wind enters the
post-SN phase prematurely. Hence, paradoxically, an increase in the
star formation rate can sometimes result in a galactic wind that
reaches a smaller extend. This happens with galaxies of masses 
$M_{\rm gal}=10^8\msun$ or less, because their shallow potential well
enables the complete removal of the ISM by the galactic wind.

For galaxies with mass above $10^{11}\msun$, the material ejected in the IGM 
always falls back onto the galaxy, no matter the value of $f_*$. Therefore,
in the case of energy-driven galactic winds,
lower-mass galaxies are more likely to be the ones responsible for
enriching the IGM and potentially perturbing the formation of nearby
galaxies. Below $10^{11}\msun$, the extent of the galactic wind and
its mass and metal content both increase with the mass of
the galaxy at constant $f_*$. With different values of $f_*$, a less
massive galaxy can sometimes produce a larger wind.

Our current model does not take into account the effect of Type Ia SNe.
These are difficult to include, because of the uncertainties on
the lifetime of the progenitors. The simulations presented in this paper
start at redshift $z=15$, and end between redshifts $z=9$ and 6.
The corresponding time periods are shorter than $1\,{\rm Gyr}$, which is
shorter than the lifetime of several Type Ia progenitors. The energy
produced by Type Ia SNe is about 20\% of the energy produced by Type II SNe
(see Fig.~10 of \citealt{benson10}). Hence, including the Type Ia SNe would
result in a slightly larger final radius for the outflow. A Type Ia SNe can
produce up to 7 times more iron than a Type II SNe (see model W7
in \citealt{nomotoetal97}), and their contribution to the iron enrichment
of the ISM become important after $1\,{\rm Gyr}$ \citep{wiersma10}. Hence,
the abundances of iron we present in this paper are underestimated.
But because of the delay, the additional iron produced
would remain in the inner parts of the galactic wind.

We have assumed a minimum value of $M_i=8\msun$ for the minimum mass of 
SNe progenitors. However, the correct value is actually quite uncertain. 
We did a few simulations with minimum masses of $6\msun$ and $10\msun$.
Our preliminary results show differences of order 10\% in the final radius
of the outflow, and of order 20-60\% in the total mass ejected, with the
largest effect occurring when $M_i$ is reduced. We intend to study this 
in more detail in the future.

To conclude, properties of galactic winds depend on the
host galaxy properties, such as the mass or star formation 
efficiency. The history of the ISM enrichment plays a determinant role
in the chemical composition and extent of the galactic wind, and therefore
its ability to enrich the IGM. The next step will consist of implementing
this galactic outflow model into large-scale cosmological simulations
of galaxy formation and the evolution of the IGM. These will be the
first simulation of this kind to include a detailed treatment of the
stellar winds and their impact on the chemical enrichment of the IGM

\section*{acknowledgments}

This research is supported by the Canada Research Chair program and
NSERC. BC is supported by the FQRNT graduate fellowship program.

\label{lastpage}

\end{document}